\documentclass[11pt]{article}
\usepackage[utf8]{inputenc}
\usepackage[T1]{fontenc}
\usepackage{graphicx}
\usepackage{grffile}
\usepackage{longtable}
\usepackage{wrapfig}
\usepackage{rotating}
\usepackage[normalem]{ulem}
\usepackage{amsmath}
\usepackage{textcomp}
\usepackage{amssymb}
\usepackage{capt-of}
\usepackage{hyperref}
\usepackage[margin=4cm]{geometry} \usepackage[version=4]{mhchem}
\usepackage{listings} \usepackage{tikz} \usepackage{caption} \usepackage{subcaption} \usepackage{float}
\usepackage{authblk} \usepackage{stackengine} \usepackage{graphicx}
\author[]{Durham Smith}
\author[]{Grigory Tikhomirov}
\affil[]{Electrical Engineering and Computer Sciences, University of California, Berkeley}
\usetikzlibrary{quotes,shapes.geometric, arrows, angles, shapes.multipart}
\definecolor{lightblue}{rgb}{0.19, 0.55, 0.91}
\date{}
\title{small: A Programmatic Nanostructure Design and Modelling Environment}
\hypersetup{
 pdfauthor={Durham Smith},
 pdftitle={small: A Programmatic Nanostructure Design and Modelling Environment},
 pdfkeywords={},
 pdfsubject={},
 pdfcreator={Emacs 29.0.50 (Org mode 9.5)}, 
 pdflang={English}}
\begin{document}

\maketitle
\begin{abstract}
Structural DNA nanotechnology has advanced to the extent that extremely complex structures can be designed. Much of this advancement has been due to the development of automated DNA design and simulation tools. Typically, the tools (e.g. NUPAK, cadnano, OxDNA) are created for a specific task. Ideally, there would be an environment that can integrate all such DNA tools, also with non-DNA tools - for example for modelling electromagnetic field along a zero-mode waveguide made of gold nanoparticles organized on a DNA breadboard. Such an environment would streamline design in DNA nanotechnology and enable applying DNA nanotechnology principles to construct high performance materials and devices from non-DNA components.

Here we present \texttt{small} a programmatic tool that is a step towards building such an environment for designing arbitrary nanostructures. In particular we showcase how \texttt{small} has been used to create an integrated computational materials  engineering (ICME) framework for DNA nanotechnology, allowing the hierarchical design, simulation and visualization of arbitrary DNA nanostructures. Furthermore we demonstrate the design and modeling of the mode profiles and band structure of hybrid DNA-nanoparticle materials through the integration of \texttt{small} with Maxwell solvers.
\end{abstract}

\section{Introduction}
\label{sec:org7305b27}
Structural DNA nanotechnology has proven to be a robust framework for the bottom-up creation of nanostructures, with applications in nanomanufacturing, nanophotonics and electronics, catalysis, computation, molecular robotics, drug delivery, bioimaging and biophysics \cite{dey21_dna_origam}.
From pioneering work by Seeman on immobile replication junctions \cite{seeman82_nucleic_acid_junct_lattic} to the introduction of DNA origami by Paul Rothemund \cite{rothemund06_foldin_dna_to_creat_nanos_shapes_patter} DNA nanotechnology has now become a field uniting hundreds of labs around the world working on constantly increasing complexity and developing new concepts. Notable approaches include single-step assembly single-stranded tiles \cite{wei12_compl_shapes_self_assem_from,ke12_three_dimen_struc_self_assem} and hierarchical assembly of DNA origami tiles \cite{tikhomirov17_fract_assem_microm_scale_dna}. This diversity in structure creation, coupled with advances in incorporating non-DNA molecules into DNA nanostructures \cite{Yang_2015,Stephanopoulos_2019,Madsen_2019}, provides the nanotechnologist with a diverse toolbox for creating nanoscale structures and systems.

The success of structural DNA nanotechnology can partly be attributed to the development of design and simulation tools \cite{dey21_dna_origam}. While there have been significant advances in computer-aided design (CAD) tools for designing DNA nanostructures, the ability of these tools to incorporate non-DNA molecules is still in its infancy \cite{glaser21_art_desig_dna_nanos_with_cad_softw,dey21_dna_origam,kekic20_in_silic_model_dna_nanos} and is limited to the visualization of few supported non-DNA molecules alongside designed DNA structures \cite{llano20_adenit} \cite{poppleton_oxview}. Furthermore the design of a nanostructure and the simulation of its properties are typically handled by different software tools. While progress has been made in integrating design and simulation tools, it has been limited to structural \cite{huang21_integ_comput_aided_engin_desig_dna_assem} and thermodynamic \cite{llano20_adenit} properties of the designed nanostructures.

Here we present \texttt{small}, a modular, extensible, programmatic integrated computational materials engineering (ICME) \cite{olson97_comput_desig_hierar_struc_mater} framework for the design and modeling of arbitrary nanostructures. Here, being extensible refers to the ability for users to create custom modules that add features to \texttt{small}, such as integrations with third party tools or models describing chemical entities. Modularity refers to the ability for users to select specific extension relevant to their particular application. We focus on using \texttt{small} to model of DNA nanostructures. In particular we showcase how arbitrary DNA nanostructures can be created in \texttt{small} and how \texttt{small} can be extended to model DNA-nanoparticle hybrid materials and simulate their electromagnetic properties. We also introduce a package manager for extensions built for \texttt{small}. This package manager enables users to easily upload designs and extensions to a repository where documentation for them is automatically generated, and provides a way to programmatically retrieve and install the extensions. The result is a tool for designing nanostructures than is flexible enough to be extended to model the unique requirements of a particular research project and allows these extension to be easily distributed to nanotechnology community.

\section{Existing Approaches}
\label{sec:org554b882}
DNA CAD tools can be grouped into three categories based on their paradigm of user interaction; graphical user interface (GUI) based, top-down or programmatic. With GUI based tools structures are created and modified by the user using the mouse to perform operations such as strand creation or ligation by pointing, clicking and dragging within a graphical window. While GUI tools are useful in that they reduce the barrier to entry for designing DNA nanostructures, they suffer from increased probability of user error as the size of DNA nanostructures grows, as well as having much higher hardware requirements than their top down or programmatic counterparts. Furthermore GUI based tools are unsuitable for tasks such as algorithmic optimization of designed nanostructures \cite{benson19_evolut_refin_dna_nanos_using} as changes to the design are required to be manually made by the user.
Top down tools take a high level structural description, typically the location of vertices in a polygon, and automatically calculate the sequence of DNA staple strands needed to assemble such a structure \cite{jun19_autom_sequen_desig_wiref_dna,jun19_autom_sequen_desig_polyh_wiref,jun21_rapid_protot_arbit_wiref_dna_origam,benson16_comput_aided_produc_scaff_dna,petrosyan2021nanoframe}. The downside of top down tools is that they typically only support the creation of wireframe DNA origami with a specific edge type and constrain the structure to certain types of polyhedra. With programmatic tools users write code to specify their design, which once run can produces files for the visualization and simulation of the resulting structure as well as the DNA sequences required to assemble the structure. Current programmatic tools are best suited to designs consisting of antiparallel helices \cite{doty20_scadn,codenano,douglas09_rapid_protot_dna_origam_shapes_with_cadnan}. In addition another common issue with programmatic DNA CAD tools, with the exception of scadnano \cite{doty20_scadn} is their incomplete, or missing, documentation, and lack of support and maintenance on the underlying codebase.

The available DNA CAD and simulation tools reveal a further issue with DNA-related software; duplication of effort. There are many common operations all DNA CAD tools would need to implement, such as placement of nucleotides and creation of staple strands.  A general software platform that provides these operations would enable users to implement their own structures or algorithmic structural design tools on top of this platform as opposed to re-implementing this functionality for every new tool. Furthermore, if this software platform was able to integrate with third party software, such as MD simulation packages, nanostructures and tools created using that platform would have access to the functionality the third party software, increasing their utility.

Coupling this with the ability to easily distribute tools, new models describing chemical entities, designed nanostructures and integrations with third party software (all of which will be referred to as \emph{extensions} hereafter) built using this platform would allow for the creation of libraries components with well understood properties. Additionally being implemented using underlying platform would allow many such components to be used as sub-components in larger designs, an essential factor for increasing the pace of innovation in the field \cite{fink2019mathematical}.

\section{An extensible platform to design nanomaterials}
\label{sec:orgce24fd6}
\texttt{small} aims to be a flexible tool for designing and modeling nanostructures and systems. To achieve this five criteria were set out when designing \texttt{small}; \emph{scalability}, \emph{extensibility}, \emph{accessibility}, \emph{good documentation} and for it to be \emph{future-proof}. In this section we detail the motivation for these criteria and how \texttt{small} achieves them.

\begin{figure}[h]
  \centering

\begin{subfigure}{.1\textwidth}
  \centering
  \includegraphics[width=.7\linewidth]{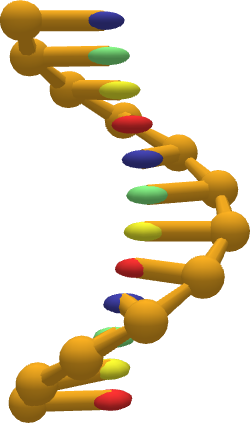}
 % \caption{}
\end{subfigure}
\hsmash{\scalebox{1.4}{$\rightarrow$}}%
% \begin{subfigure}{.59\textwidth}
%   \centering
%   \includegraphics[width=.7\linewidth]{./img/dna-hierarchy}
%   \caption{}
% \end{subfigure}
 \begin{subfigure}{.20\textwidth}
   \centering
   \includegraphics[width=.7\linewidth]{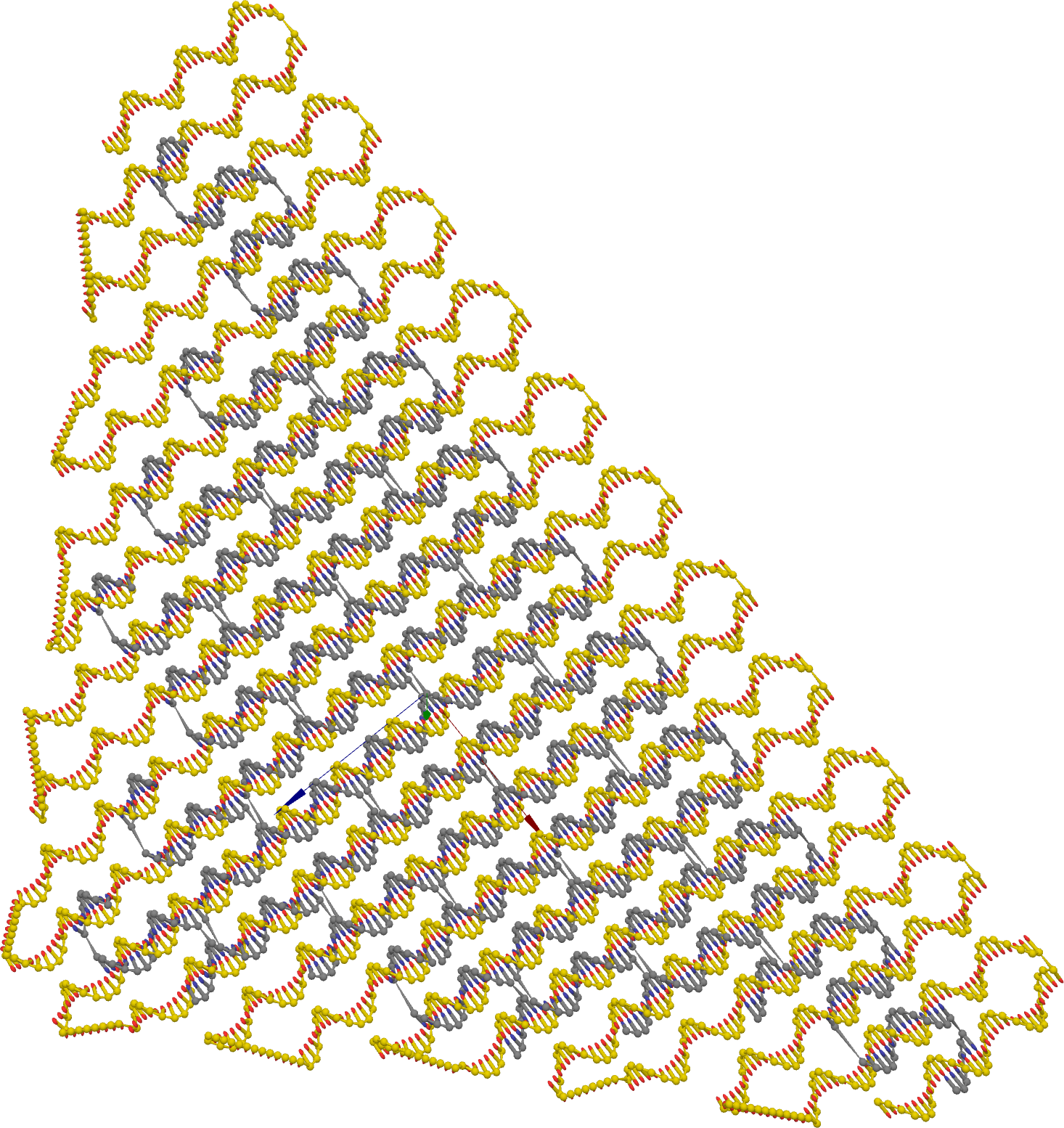}
   %\caption{}
   \label{fig:hierachy-triangle}
 \end{subfigure}%
 \hsmash{\scalebox{1.4}{$\rightarrow$}}%
\begin{subfigure}{.20\textwidth}
  \centering
  \includegraphics[width=.7\linewidth]{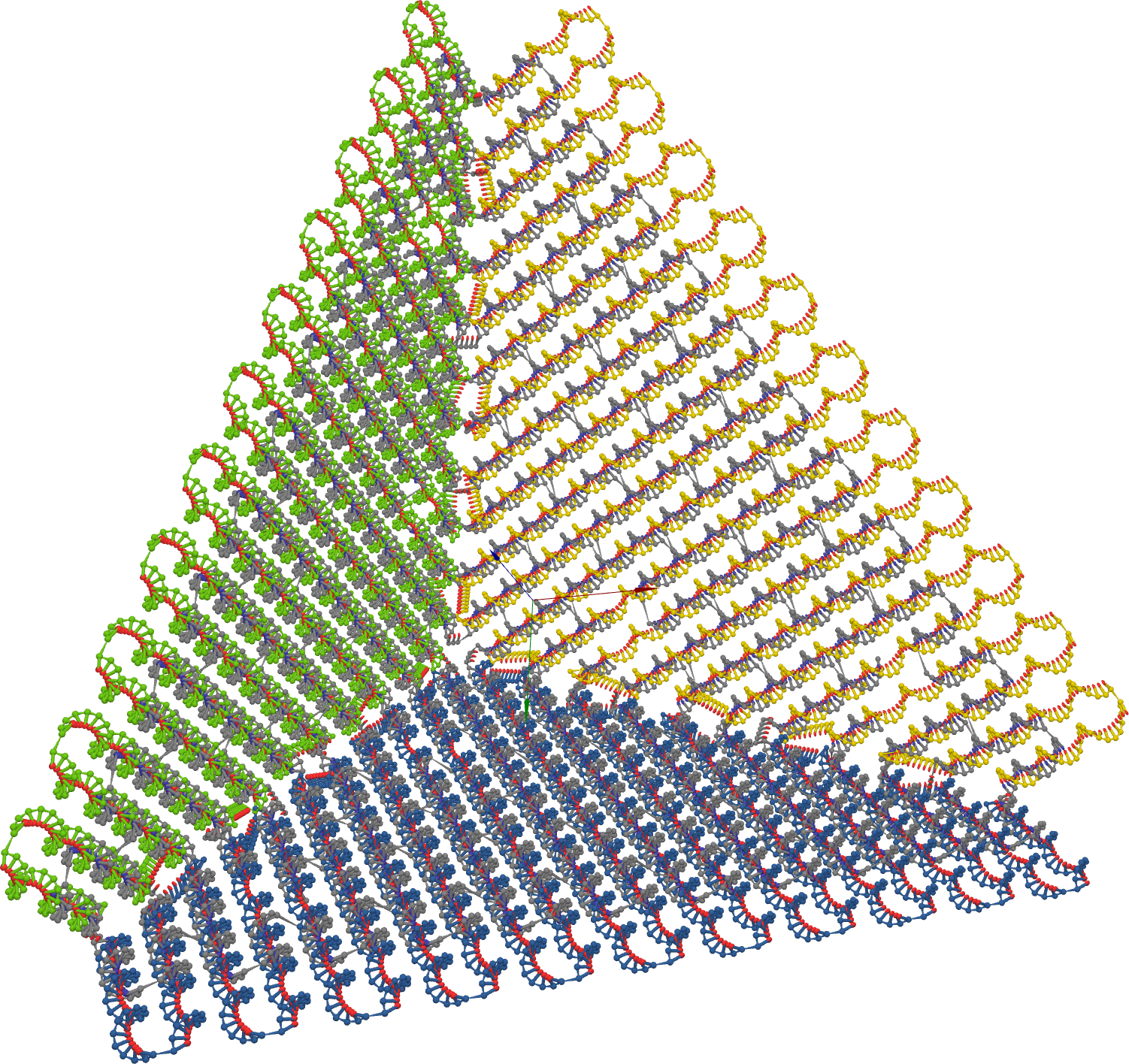}
  %\caption{}
  \label{fig:hierarchy-cone}
\end{subfigure}
\hsmash{\scalebox{1.4}{$\rightarrow$}}%
\begin{subfigure}{.20\textwidth}
  \centering
  \includegraphics[width=.7\linewidth]{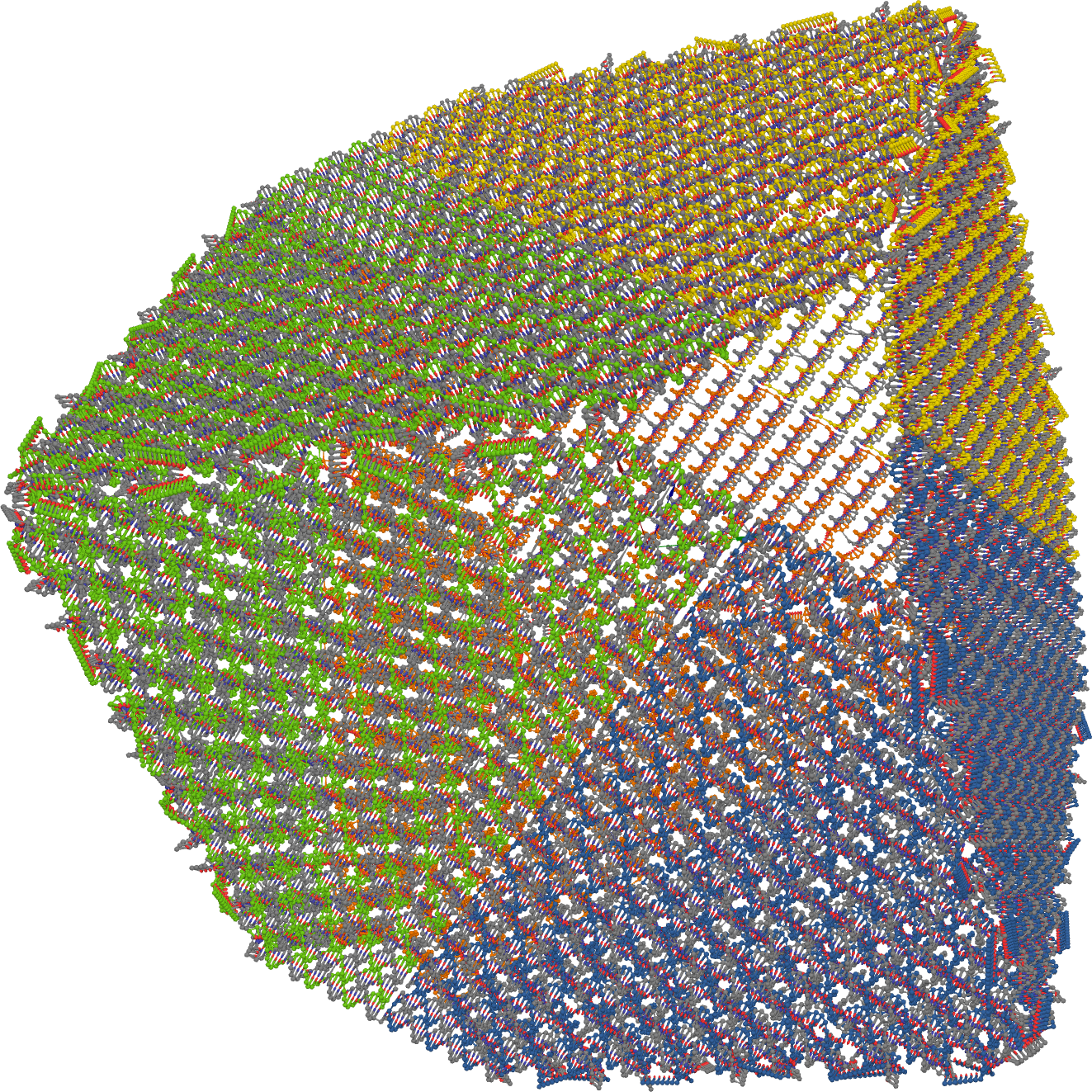}
  %\caption{}
  \label{fig:hierarchy-cube}
\end{subfigure}
\hsmash{\scalebox{1.4}{$\rightarrow$}}%
\begin{subfigure}{.25\textwidth}
  \centering
  \includegraphics[width=.7\linewidth]{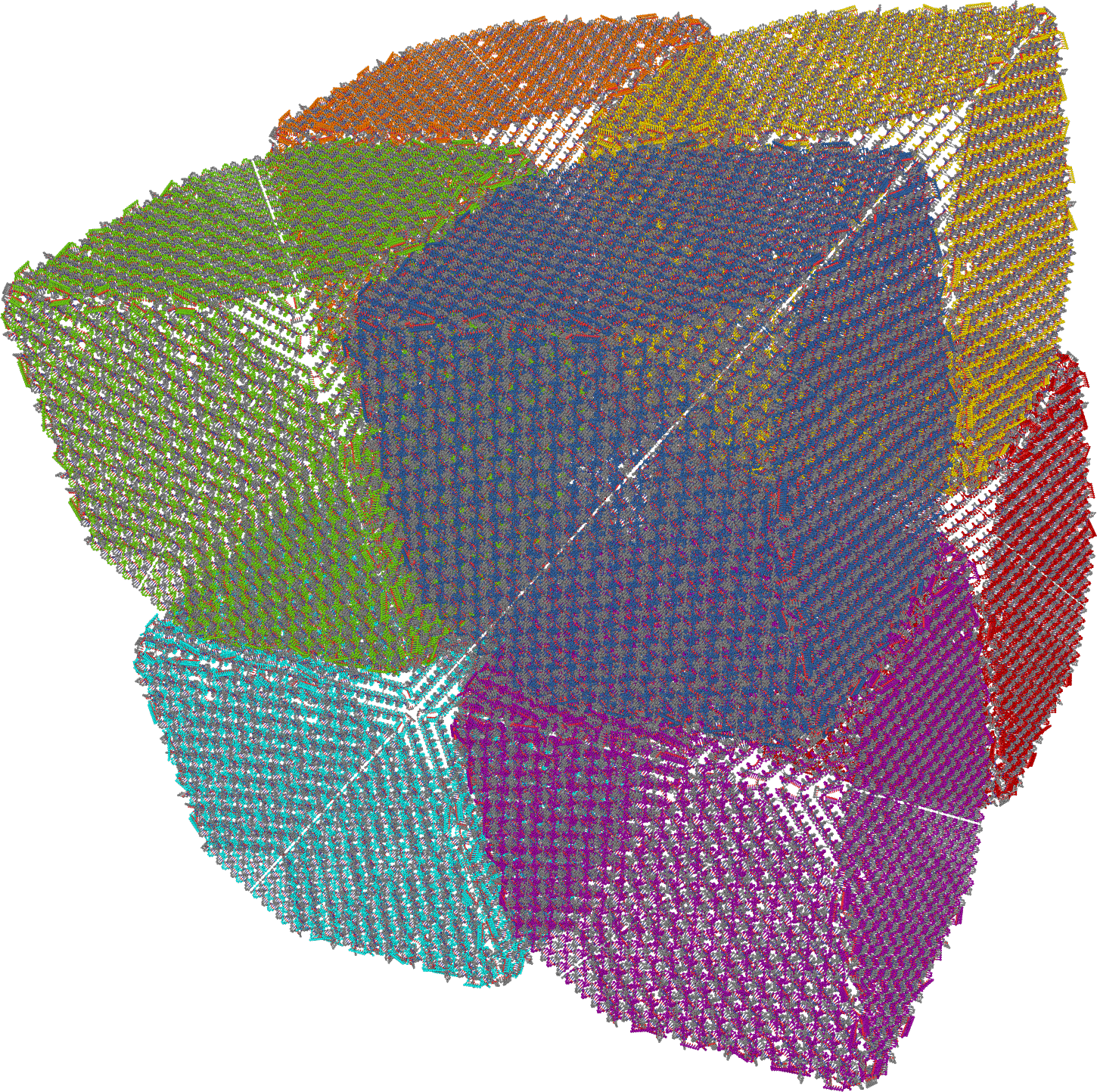}
%  \caption{}
  \label{fig:hierarchy-nt}
\end{subfigure}%
\caption{Hierarchical assembly of a \(2\times 2\times 2\) array of DNA cubes. Each successive structure is created from the hierarchical assembly of multiple identical components from the previous step. Colors indicate individual hierarchical units from the previous assembly step. Figures rendered in oxView using the oxView extension.}
% {(a) An example of the class hierarchy \lstinline{small} describing DNA origami. Classes describing DNA entities are given as a box split horizontally. The top section of the box gives the class name in blue (dark blue for the actual class and light blue for its superclasses) and arrow point to the direct superclass. (b) A single triangle from the tile of Tikhomirov et al \cite{tikhomirov17_fract_assem_microm_scale_dna}, with an examples of DNA-NT, DNA-SINGLE-STRAND, DNA-HELIX-STRAND and DNA-STAPLE-STRAND indicated in red, dark blue, light blue and green respectively. The full triangle is example of the DNA-ORIGAMI class. Note that the geometry of the DNA-SINGLE-STRAND's nucleotides are different geometric to that of DNA-HELIX-STRAND, demonstrating the fine grained level of control over structures built in small
% }
\label{fig:hierarchy}
\end{figure}

Scalability in this context refers to an ability to model nanosystems on any length scale. This is important since nanoscale systems and materials can be combined to form larger systems or materials, for example in the fractal assembly of DNA origami arrays \cite{tikhomirov17_fract_assem_microm_scale_dna} or in nanoparticle assemblies \cite{kim20_mie_reson_three_dimen_metac,rogers16_using_dna_to_progr_self}. \texttt{small} achieves scalability in two ways. First, \texttt{small} provides methods for arbitrary describing chemical entities, which is done through inheritance of the \texttt{CHEM-OBJ} class. This class provides the machinery for performing geometric transformations on chemical entities but does not enforce the use of a particular model for their description. Describing a specific entity requires implementing a model that provides a coordinate description for key features of the chemical entity. For example atoms could be represented by \(x,y,z\) coordinates, whereas nucleotides might use a model similar to that in figure \ref{fig:nt-model}. The second way \texttt{small} achieves scalability is by allowing the hierarchical composition of chemical entities. Again the mechanisms to do so are provided by the \texttt{CHEM-OBJ} base class. This allows larger chemical entities, termed \texttt{parents}, to be created from multiple smaller ones, termed \texttt{children}. In turn these hierarchical chemical entities can be used as individual components in a design or as \texttt{child}ren in the creation of larger entities. Supplementary video \href{https://youtu.be/NqbKlmdHW8g}{1} and \href{https://youtu.be/h00R_PvjieM}{2} provide an introduction to \texttt{Common Lisp} and \texttt{small}, focusing on the necessary concepts for describing arbitrarily chemical entities by defining a model for them and grouping them using the parent-child hierarchy. Supplementary video \href{https://youtu.be/nQVsvWV6eUk}{3} shows how these concepts have been used to model DNA at the nucleotide, strand and origami level and supplementary video \href{https://youtu.be/KiAtNQvJ3SY}{6} provides an example of the hierarchical assembly of an array of DNA cubes, seen in figure \ref{fig:hierarchy}.

\begin{figure}[h!]
  \centering
 \begin{subfigure}[t]{.35\textwidth}
   \centering
   \includegraphics[width=.95\linewidth]{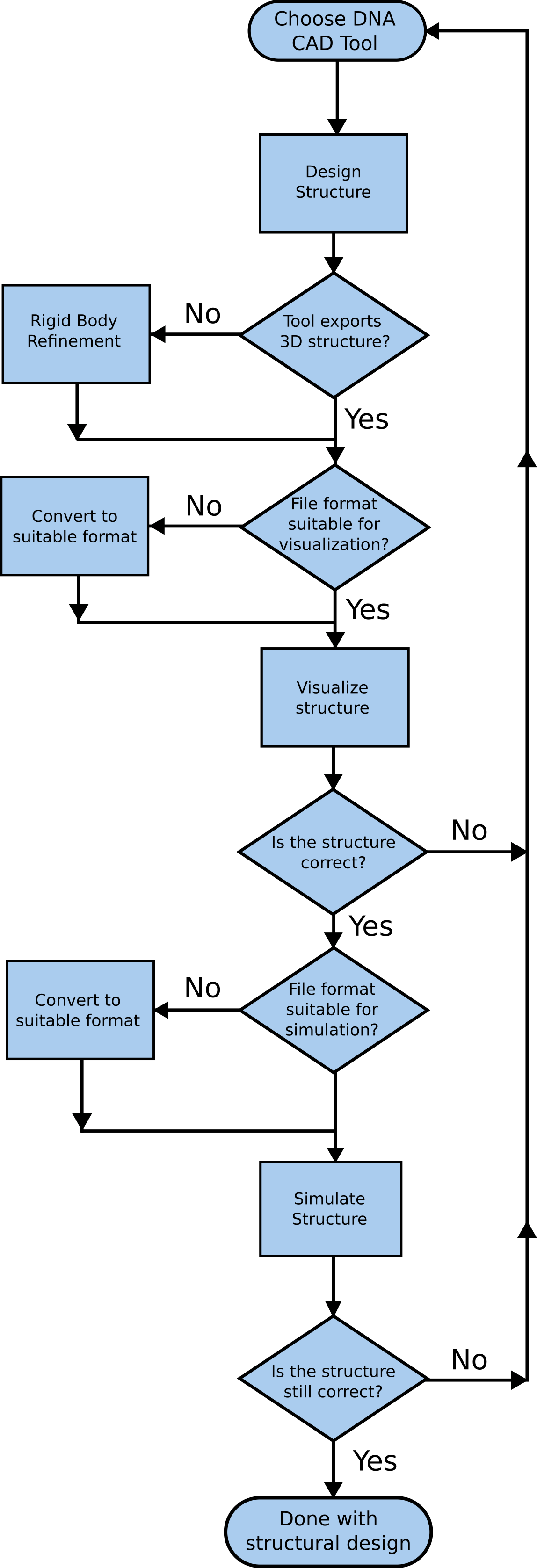}
   \caption{}
   \label{fig:workflow-trad1}
 \end{subfigure}%
\begin{subfigure}[t]{.29\textwidth}
  \centering
  \includegraphics[width=.95\linewidth]{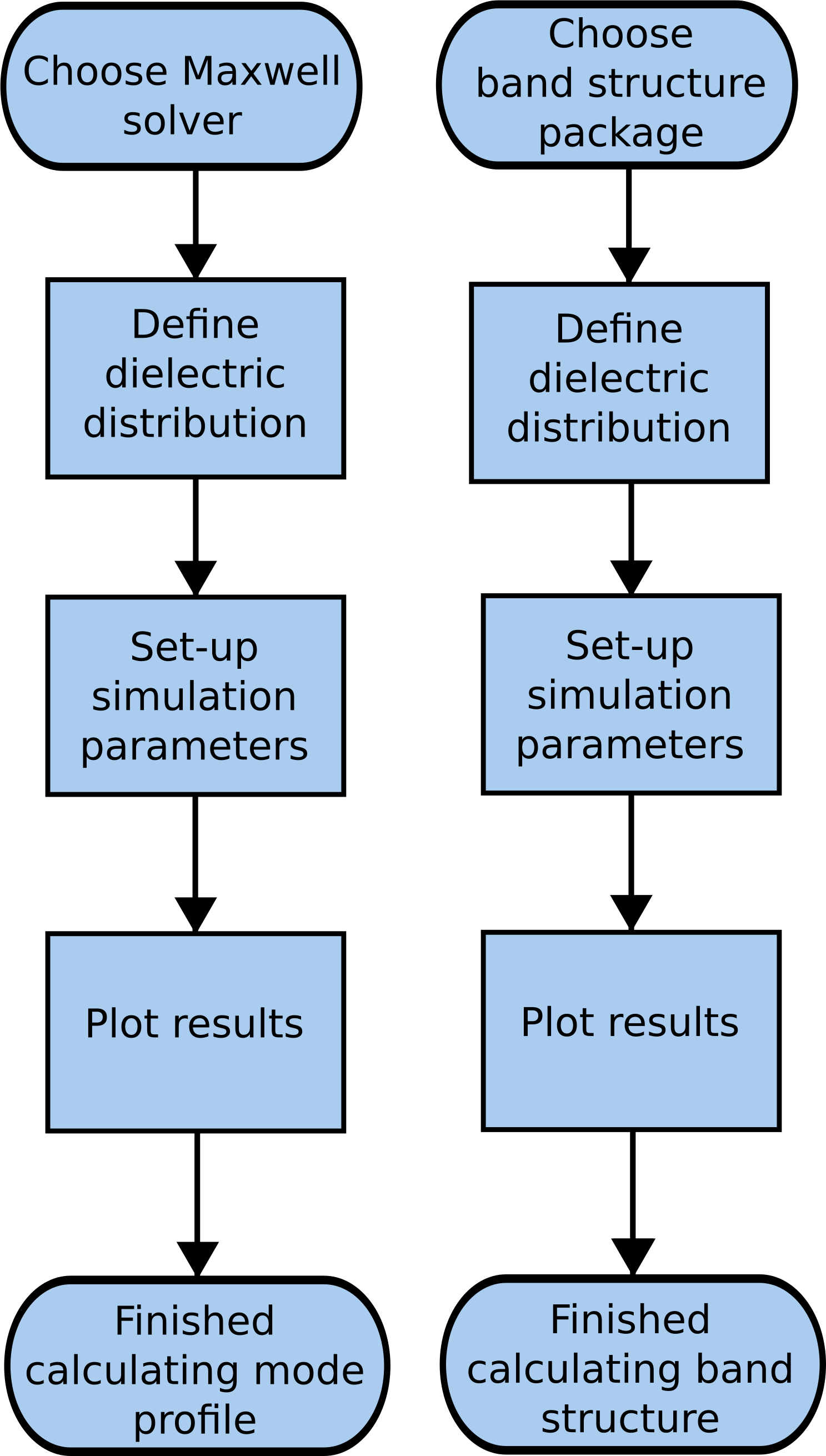}
  \caption{}
  \label{fig:workflow-trad2}
   \end{subfigure}%
\begin{subfigure}[t]{.35\textwidth}
  \centering
  \includegraphics[width=.95\linewidth]{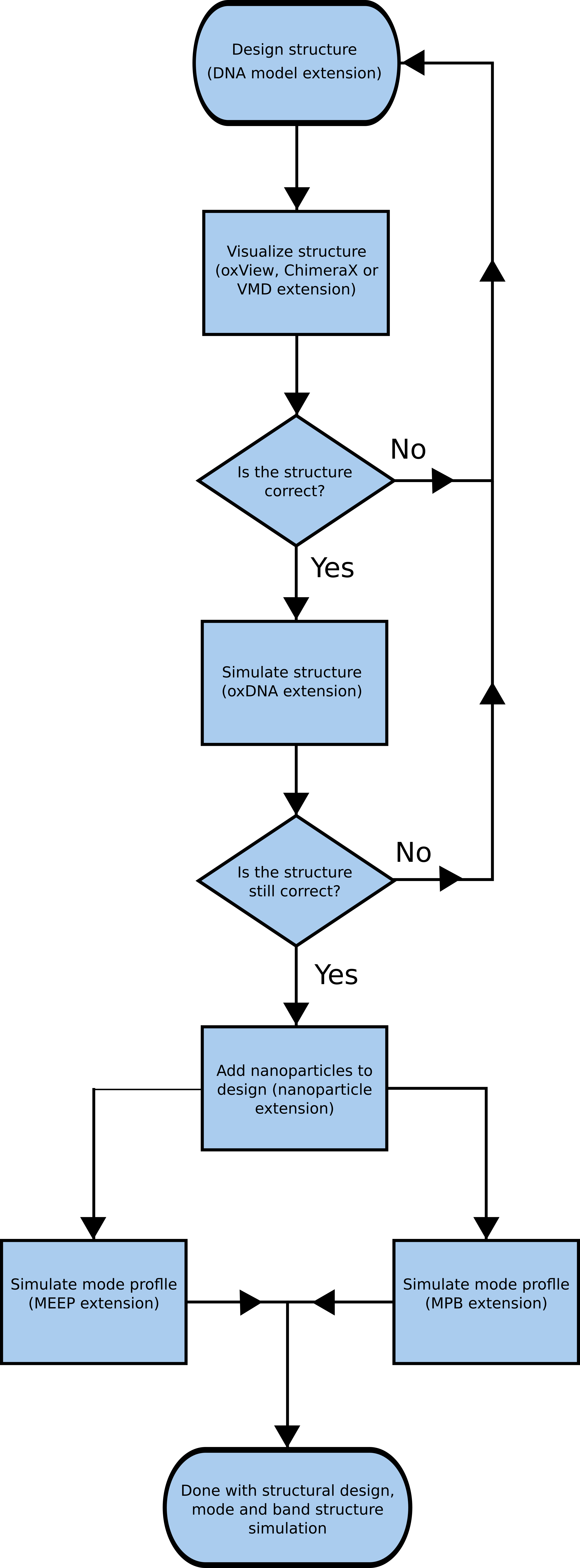}
  \caption{}
  \label{fig:workflow-small}
\end{subfigure}%

\caption{Example of workflows for the design and simulation of array of cubic nanoparticles encased in DNA origami using traditional methods, a and b, and \lstinline{small}, c. (a) Traditional workflow for DNA structure creation, note each step may require use of a different software package. (b) Traditional workflow for simulating the mode profile (left) and band structure (right), note that while the steps are similar they are performed in different software packages and information from the designed DNA nanostructure is not used explicitly in setting up these simulations. (c) Workflow using \lstinline{small}, note how all required software packges can be accessed through extensions to \lstinline{small} and information from the DNA nanostructure is used setting up simulations of the mode profile and band structure of the array.}
\label{fig:workflows}
\end{figure}

The second design principle is \emph{extensibility}, which refers to an ability to add features, such as integration with third party tools, to \texttt{small}. As nanostructures and systems become more complex and are used to create systems and materials with desired micro or macroscopic properties, multiple tools are needed for their design and analysis. Typically, these tools are released as stand-alone software packages, requiring individual researchers crafting workflows suited their particular research project. An example of such a workflow for designing an hybrid array of cubic nanoparticles encased in DNA and simulating its mode profile and band structure can be seen in figures \ref{fig:workflow-trad1} and \ref{fig:workflow-trad2} . Allowing these tools to be directly integrated with the design software, removes the need for developing independent workflows for these tasks and also allows information from each phase of the design to be used in subsequent phases, as seen in figure  \ref{fig:workflow-small}. Examples of integrating \texttt{small} with third party software can be seen in supplementary video \href{https://youtu.be/3zEAnOo9BIo}{4}, which gives the integration with visualization software packages oxView, ChimeraX and VMD, supplementary video \href{https://youtu.be/3zEAnOo9BIo}{5} showing the integration with the oxDNA molecular dynamics simulation package, and supplementary video \href{https://youtu.be/3zEAnOo9BIo}{7} detailing the integration MIT Photonic Bands (MPB) package respectively.

The third design principle is \emph{accessibility}. This refers to the ability of users to freely obtain the software, extensions to it, designs of nanostructures made using it, and also to the ability for the software to be run on an operating system of the users choice. Closely related to accessibility is the fourth design principle, \emph{good documentation}. For \texttt{small}, designs of nanostructures created in it, and extensions to it, to be easily usable and quickly adopted by researchers it is required that their usage be well documented. The last design principle is that \texttt{small} should be \emph{future-proof}. This means that designs and extensions written in \texttt{small}, and \texttt{small} itself, should not 'break' over time, for example when dependencies change their API or newer versions of underlying language that \texttt{small} is built on are released. These design goals are in part supported by the language that \texttt{small} is written in, \texttt{Common Lisp} and in part through the addition of a repository for extensions and package manager. \texttt{Common Lisp} has a well defined standard and many freely available conforming implementations for all major operating systems and architectures. Furthermore \texttt{Common Lisp} supports adding documentation to code and the ability to interface with code bases written in other languages. These features are extensively used by \texttt{small}'s repository, figure \ref{fig:repo}, and package manager which facilitate the distribution of user defined extensions to \texttt{small} and its use is detailed in supplementary video \href{https://youtu.be/8v9u-20fAwI}{8}.

\section{A ICME for DNA nanotechnology}
\label{sec:org298c23d}
In this section we provide an example of \texttt{small}'s ability to model arbitrary chemical entities and integrate with different software tools. We focus on DNA nanotechnology and the use of \texttt{small}'s parent-child hierarchy to create descriptions of DNA on the nucleotide, strand and origami level. Furthermore we showcase the integration with various molecular visualization software and the molecular dynamics simulation package oxDNA. The result is an integrated computational materials engineering framework for DNA nanotechnology.

\begin{figure}[h]
    \centering
    \begin{subfigure}[b]{0.45\textwidth}
        \centering
        \begin{tikzpicture}
          \tikzstyle{cm_circf} = [circle,fill=blue!50, minimum size=0.5cm]
          \tikzstyle{cm_circb} = [circle,fill=blue!20, minimum size=0.5cm]
          \tikzstyle{hel_elipse} = [ellipse, minimum height=4cm, draw=black, dashed]
          \node[hel_elipse] at (0,0) {};
          \node[hel_elipse] at (4,0) {} ;
          \draw [dashed] (0,0) -- node[above] {} (4,0);
          \draw [thick] (0,2) -- node[above] {} (4,2);
          \draw [thick] (0,-2) -- node[above] {} (4,-2);
          \node[cm_circb] at (3.6,0.8) {};
          \node[cm_circb] at (2.8,1.16) {};
          \node[cm_circb] at (2,1.12) {};
          \node (nt1) [cm_circb] at (1.2,0.68) {};
          \node (nt0) [cm_circf] at (0.4,0) {};
          % \draw [gray, dashed, label=below:{\(2\))}] (0,0) circle [radius=2cm] {Hello1};
          % \draw [fill=black] (-1.2,0) circle [radius=0.1cm];
          \draw [fill=black] (0,0) circle [radius=0.01cm];
          \draw [|-|] (1.2,1.5) -- node[above] {\lstinline{*helix-nt-spacing*}} (2,1.5);
          \draw [|->, thick] (.4,0) -- node[above] {\lstinline{vn}} (1.2,0);
        \end{tikzpicture}
        %\caption{$y=x$}
        %\label{fig:y equals x}
      \end{subfigure}
      \hfill
      \begin{subfigure}[b]{0.45\textwidth}
        \centering
        \begin{tikzpicture}
          \tikzstyle{cm_circf} = [circle,fill=blue!50, minimum size=0.5cm]
          \tikzstyle{cm_circb} = [circle,fill=blue!20, minimum size=0.5cm]
          \tikzstyle{hel_circ} = [circle, minimum size=4cm, draw=black, dashed]
          \tikzstyle{origin} = [circle, minimum size=0.001cm, fill=black]
          \node (origin) [hel_circ] at (0,0) {};
          \node (cm) [cm_circf] at (-1.2,0) {};
          \node (cm2) [cm_circb] at (-1.01,0.68) {};
          % \draw [gray, dashed, label=below:{\(2\))}] (0,0) circle [radius=2cm] {Hello1};
          % \draw [fill=black] (-1.2,0) circle [radius=0.1cm];
          % \draw [fill=black] (0,0) circle [radius=0.01cm];
          \draw [dashed]
          (0,0) coordinate (a) node[right] {}
          -- (-1.648593,1.13231) coordinate (b) node[left] {\lstinline{*helix-radius*}};
          % \draw [|-|] (0,0) -- node[ left] {\lstinline{*helix-radius*}} (-1.648593,1.13231);
          \draw [|-|] (-1.2,-.7) -- node[below, xshift=0.6cm] {\lstinline{*helix-cm-offset*}} (0,-.7);
          \draw [->|, thick] (-1.2,0) -- node[below] {\lstinline{vbb}} (0,0);
          % \draw (0,0) arc (0:34.48:1cm);
          % angle
          \pic ["$\alpha$", draw=blue, <->, angle radius=8mm, angle eccentricity=1.2] {angle = cm2--origin--cm}  ;
          \node[right] at (-.4,0.5) {$\alpha$ \lstinline{= *rad/bp*}};
        \end{tikzpicture}
        %\caption{$y=3sinx$}
      \end{subfigure}
      \caption{DNA model implemented by \lstinline{small}. Left: projection parallel to the helical axis, Right: projection perpendicular to the helical axis. Centers of mass of the nucleotides and the parameters that govern the model are indicated in blue.}
      \label{fig:nt-model}
\end{figure}
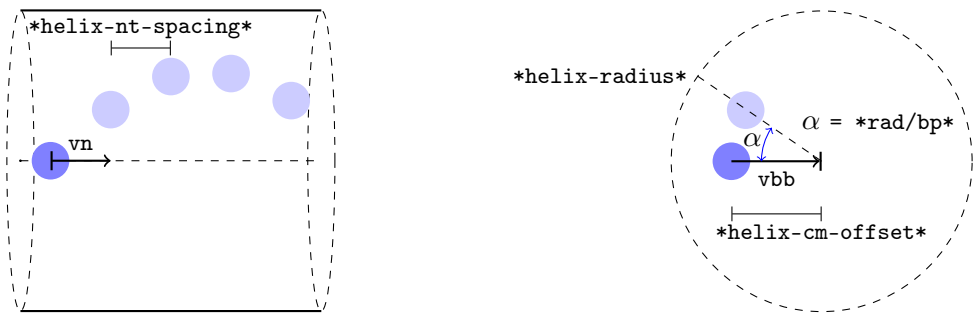

The geometric model for DNA nucleotides is implemented by the \texttt{DNA-NT} class, which inherits from \texttt{CHEM-OBJ}, can be seen in figure \ref{fig:nt-model}. A nucleotide's position is described by three vectors; the vector of the center of mass \(\mathbf{v}_{cm}\) in Cartesian coordinates, a vector normal to the face of the nucleotide's base in the \(5^{\prime}\rightarrow 3^{\prime}\) direction, \(\mathbf{v}_{n}\), and a vector directed from the nucleotide's center of mass towards its backbone, \(\mathbf{v}_{bb}\). Spacing and angles between successive nucleotides in a DNA strand are governed by the variables \texttt{*helix-nt-spacing*} and \texttt{*helix-nt-spacing*} and can easily be modified. This is a unique feature for DNA design tools and is useful for modeling situations where helices are placed under strain \cite{dietz09_foldin_dna_into_twist_curved_nanos_shapes}.

The model for DNA strands is implemented by the \texttt{DNA-STRAND} class. This class is an example of using the parent-child hierarchy to describe larger chemical entities composed of smaller ones. A \texttt{DNA-STRAND} contains many children \texttt{DNA-NT}s. Similarly \texttt{small} provides a \texttt{DNA-ORIGAMI} class to model DNA origami which contains many \texttt{DNA-STRAND}s as it's \texttt{child}ren. These \texttt{DNA-ORIGAMI} objects can themselves be use in as \texttt{child}ren in larger designs, allowing the hierarchical composition of arbitrarily complex DNA structures.

\href{https://youtu.be/nQVsvWV6eUk}{Supplementary video 3} gives an introduction to \texttt{small}'s DNA model, and details useful functions for creating and manipulating DNA structures at the nucleotide, strand and origami level. Furthermore \href{https://youtu.be/KiAtNQvJ3SY}{supplementary video 6} details the creation of an array of DNA cubes, figure \ref{fig:hierarchy} where the each structure is the result of the hierarchical assembly of multiple structures from previous assembly steps.

It is worth noting that while these examples show the creation of DNA starting from the nucleotide level and using the \texttt{parent}-\texttt{child} hierarchy to describe and organize objects on increasing length scales, the same concepts can be used to describe objects on decreasing length scales. For example if an atomic model of DNA is required a \texttt{CHEM-OBJ} implementing a model of atoms could be defined and used as \texttt{child}ren to \texttt{DNA-NT}s.

In addition to a model for DNA \texttt{small} integrates with the visualization software oxView, ChimeraX and VMD for the visualization of designed nanostructures. Furthermore simulations of designed DNA nanostructures structures is facilitated through the integration of the commonly used oxDNA molecular dynamics simulation package. Integration of visualization software and oxDNA from within \texttt{small} is detailed in section \ref{integration} and supplementary videos \href{https://youtu.be/3zEAnOo9BIo}{4} and \href{https://youtu.be/3zEAnOo9BIo}{5} provide tutorials on their use.

\section{Modeling of multi-material nanostructures}
\label{sec:org5f1551a}
\begin{figure}[H]
    \centering
    \begin{subfigure}[b]{0.45\textwidth}
        \centering
        \includegraphics[width=1.0\textwidth]{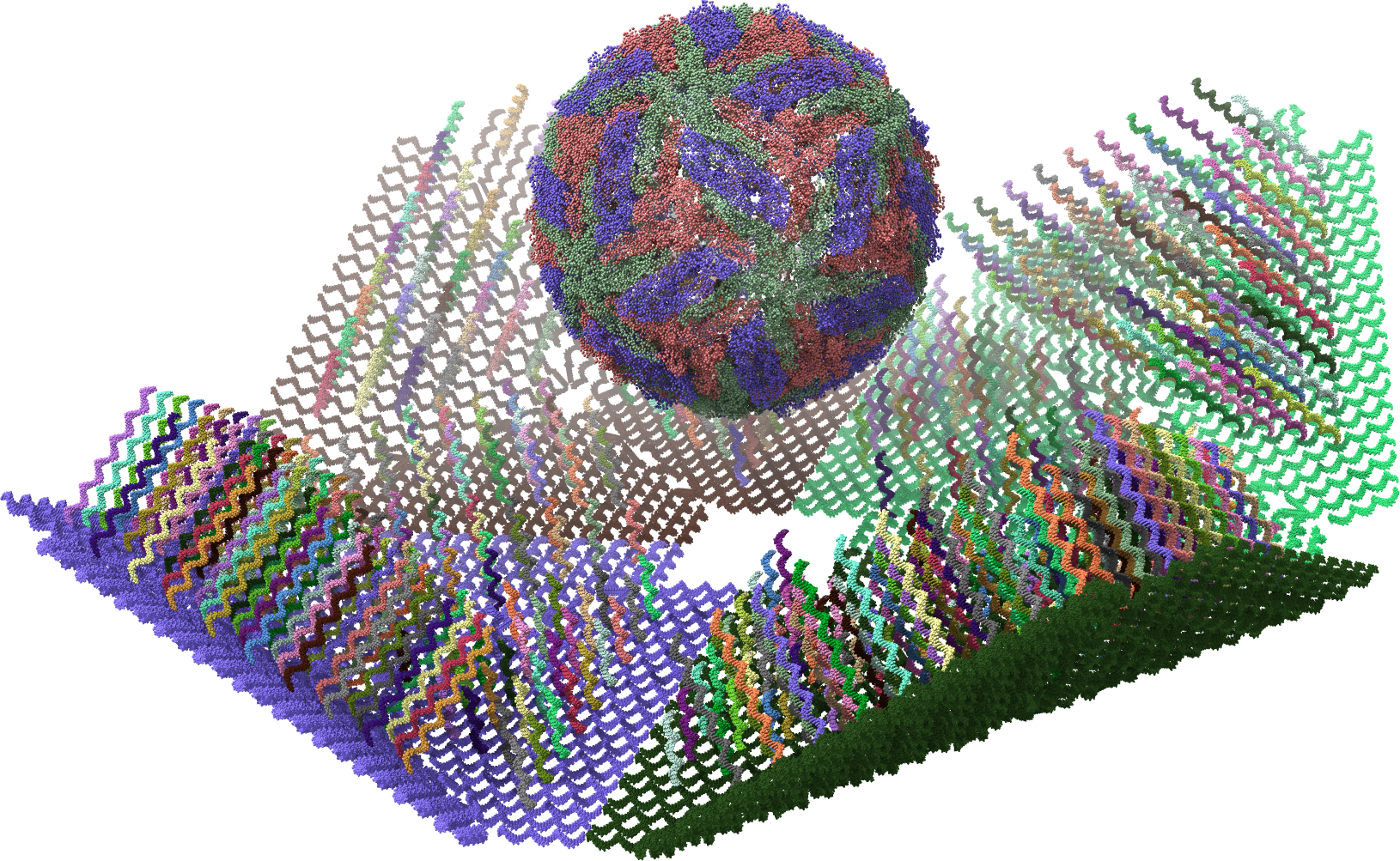}
        \caption{}
        \label{fig:virus}
      \end{subfigure}
      \hfill
      \begin{subfigure}[b]{0.45\textwidth}
        \centering
        \begin{center}
         \includegraphics[width=0.60\textwidth]{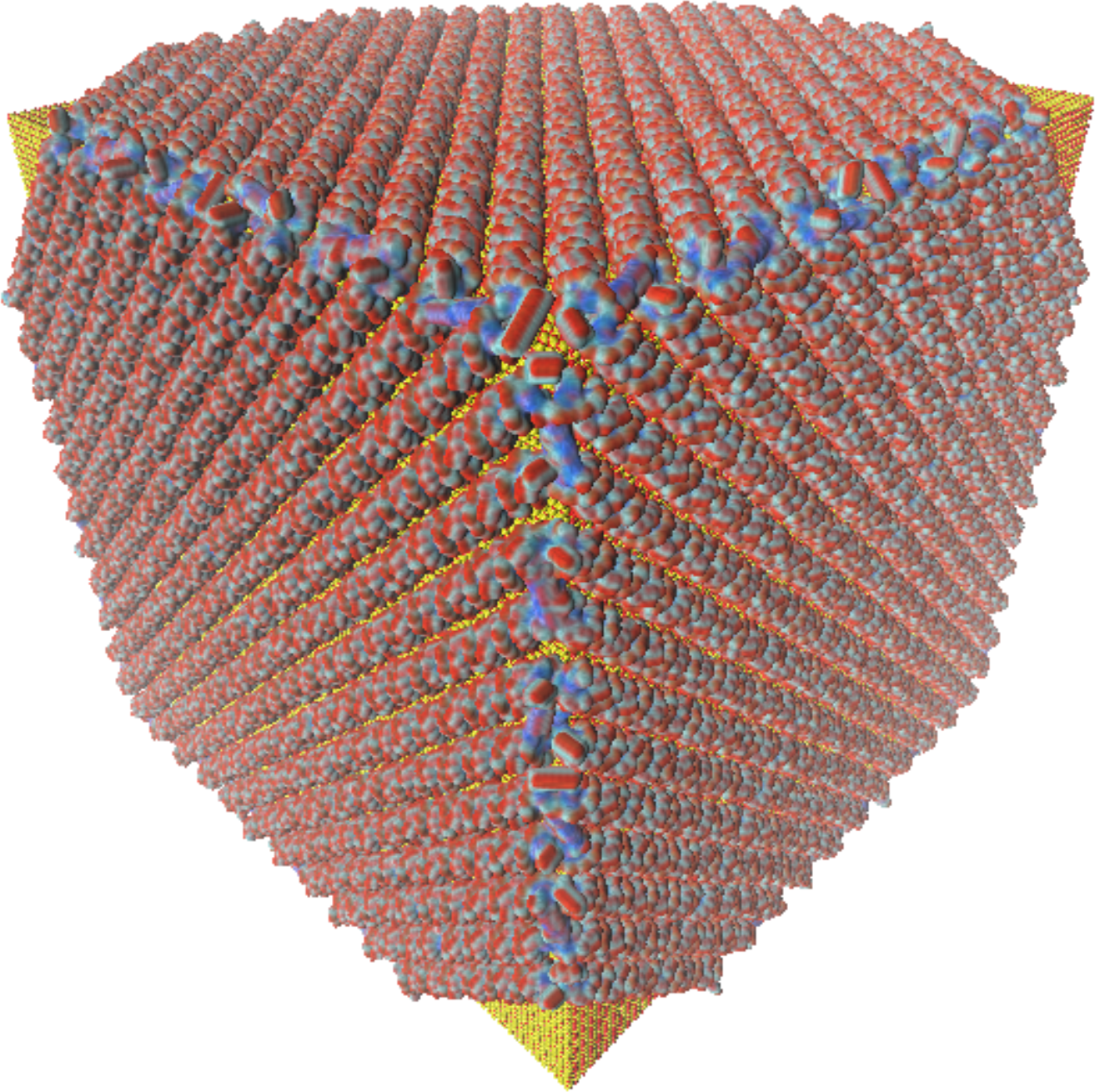}
        \end{center}
        \caption{}
        \label{fig:hybrid}
      \end{subfigure}
        \caption{(a) DNA tiles with single straded extensions aptamer extensions together with dengue fever virus. Figure rendered in ChimeraX using the ChimeraX extension. (b) \ce{SiO2} nanoparticle encased in DNA origami cube. Figures rendered in VMD using the VMD extension.}
      \label{fig:multi}
\end{figure}

An ability to model various nanoscale materials that form nanosystems or composite materials is needed if the field of nanotechnology is to live up to its expected potential.  Multimaterial modeling is still in its infancy when it comes to the field of DNA nanotechnology, and nanotechnology more broadly. Although simple geometric modeling has borne fruit, for example modeling DNA-Protein complexes has allowed for the creation of virus ligands \cite{kwon19_desig_dna_archit_offer_precis}, figure \ref{fig:virus}, more complex nanosystems and materials may require taking advantage of material properties which need to be modeled at different length scales. For example, DNA's programmability may be used to direct nanoparticles to assemble into arrays, where placement of individual nanoparticles effects the optical properties of the entire array \cite{giovampaola14_digit_metam}. In such a situation, one would need to model the DNA structures to ensure geometric shape is formed and obtain the DNA sequences needed to drive the self assembly process. The modeling of the materials optical properties would then be done by providing a description of dielectric distribution of the material to a software package that solves the Maxwell equations.

Multimaterial systems can be modeled \texttt{small} by combining models of different materials within one design. This combination of different models of chemical entities is allowed since all models of chemical entities inherent from the \texttt{CHEM-OBJ} class, which provides the functionality for combining and manipulating different models of chemical entities under a consistent API. As an example in addition to the model for DNA for model for cubic nanoparticles has also been implemented. Using these models a hybrid structure consisting of a cubic \ce{SiO2} nanoparticle encapsulated in a DNA cube made from four individual DNA origami scaffold strands can be seen in figure \ref{fig:hybrid}.

\section{Integration with nano and macro scale simulation tools}
\label{integration}
The ability to hierarchically model multi-material nanostructures allows the creation of structures with emergent properties beyond those of it's constituent components. Effective nanosystem design software should allow the modeling of both the individual components and the overall system. In this section we detail how \texttt{small} can be extended to incorporate external software packages for the simulation of different properties of the system being modeled.

\texttt{Common Lisp} provides many ways to work with external software written in other languages. The first of these is through a foreign function interface (FFI) which allows users to specify how Lisp objects should be passed to functions in the other language, and how the values the function returns should be translated back into a Lisp object \cite{weitz16_common_lisp_recip}. Another approach is to launch the program as a separate process and to use streams to communicate with that process from Common Lisp. Using these approaches \texttt{small} has integrated the functionality of micro and macro scale simulation software, molecular visualization software and chemical file format conversion tools.

\begin{figure}[H]
    \centering
    \begin{subfigure}[b]{0.14\textwidth}
        \centering
        \includegraphics[width=\textwidth]{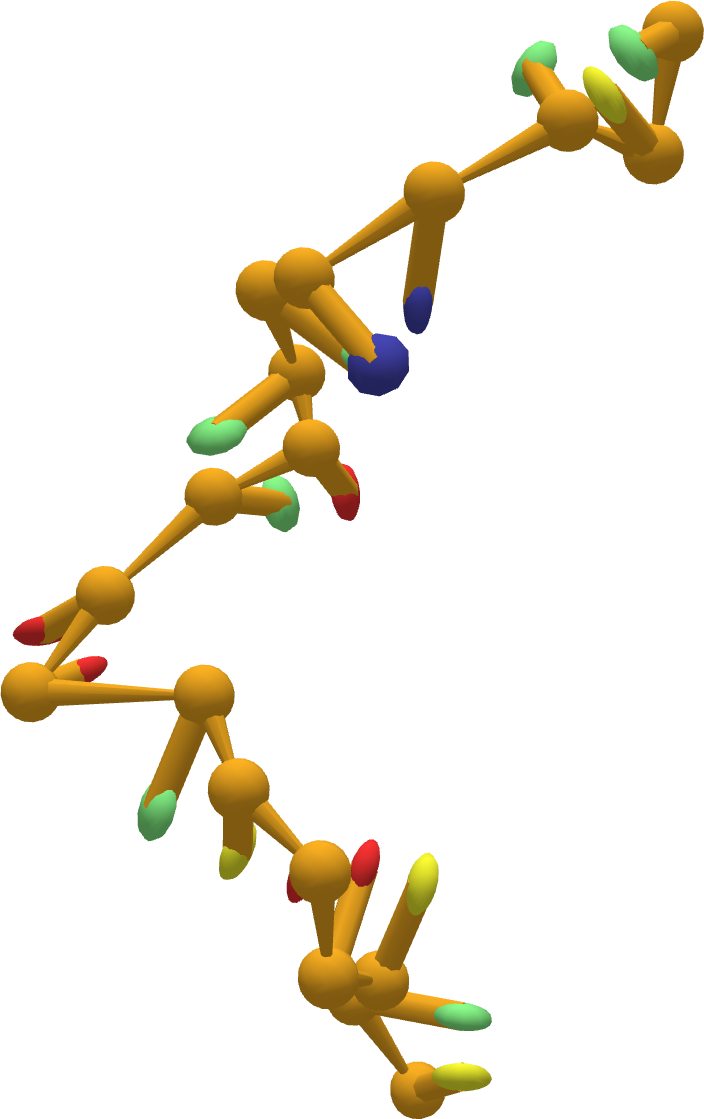}
        \caption{}
    \end{subfigure}
    \begin{subfigure}[b]{0.14\textwidth}
        \centering
        \includegraphics[width=\textwidth]{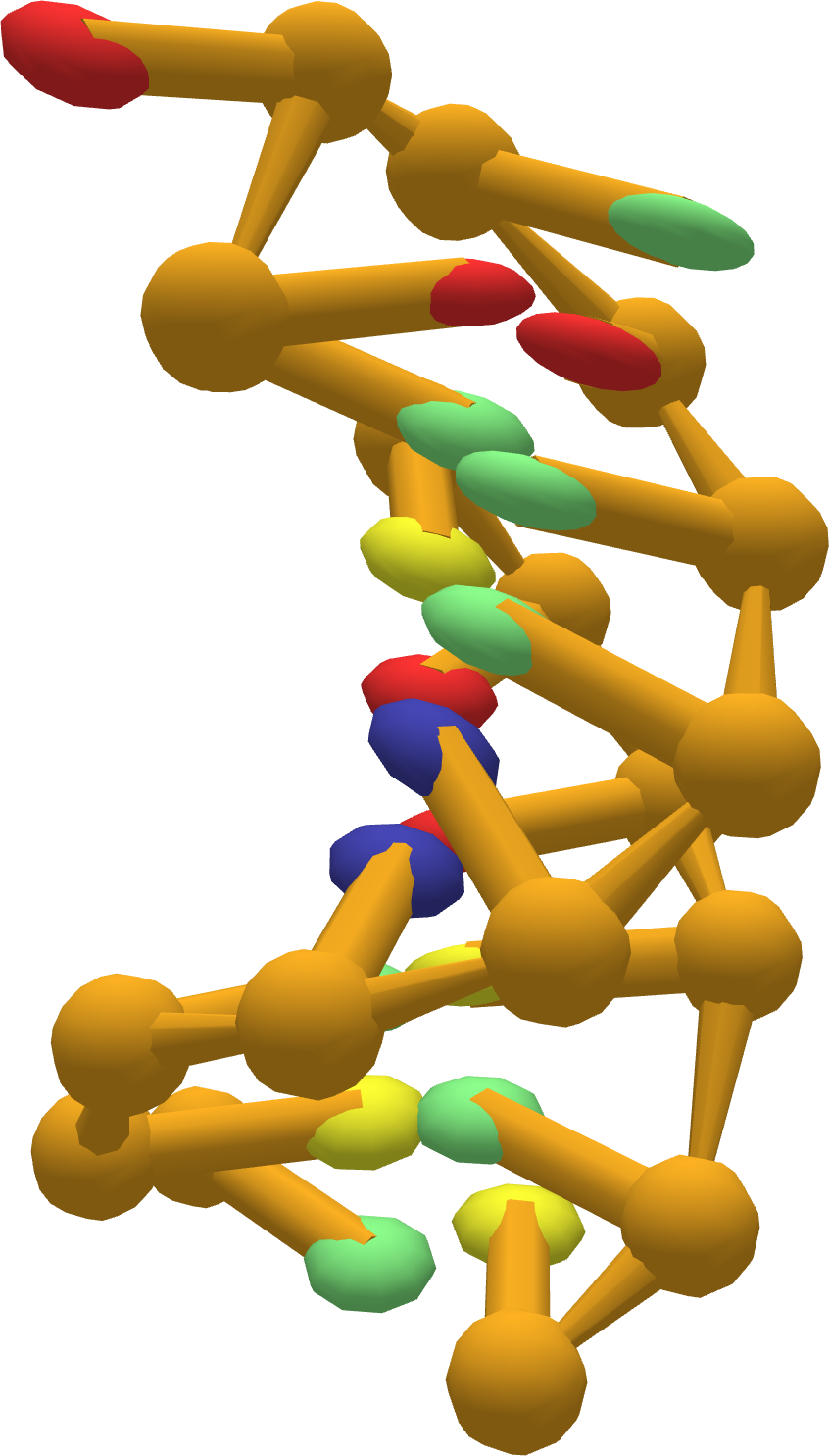}
        \caption{}
    \end{subfigure}
    \begin{subfigure}[b]{0.345\textwidth}
        \centering
        \includegraphics[width=\textwidth]{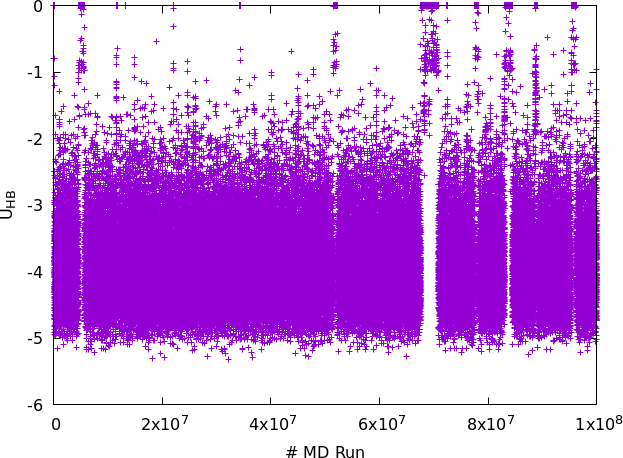}
        \caption{}
    \end{subfigure}%
    \begin{subfigure}[b]{0.345\textwidth}
        \centering
        \includegraphics[width=\textwidth]{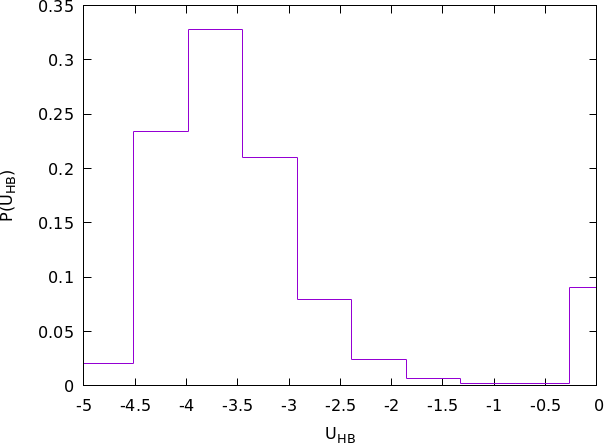}
        \caption{}
    \end{subfigure}
       \caption{oxDNA Molecular Dynamics for \(10^{8}\) molecular dynamics run at 334k on a 18 nucleotide single stranded DNA able to form a hairpin turn. (a) Initial configuration (b) Sampled low energy configuration (c) Scatter plot of hydrogen bond potential energy of MD runs (d) Histogram of hydrogen bond potential energy of MD runs.}
       \label{fig:oxdna}
\end{figure}

For microscale simulation \texttt{small} incorporates the DNA molecular dynamics package oxDNA \cite{snodin15_introd_improv_struc_proper_salt}, allowing simulations to be run on DNA structures created in \texttt{small}. Figure \ref{fig:oxdna} and \href{https://youtu.be/3zEAnOo9BIo}{supplementary video 5} shows an example of using oxDNA to simulate a 18 nucleotide DNA single strand which forms a hairpin turn. While \texttt{small} is a programmatic tool the visualization of the resulting molecular structures can be incredibly useful in assessing its correctness. For this \texttt{small} integrates with the visualization software \texttt{oxView} \cite{poppleton_oxview}, \texttt{VMD} \cite{humphrey96_vmd} and \texttt{ChimeraX} \cite{chimerax}. Furthermore different software may expect a specific file type. \texttt{small} provides by integration between common chemical file formats by integrating with the \texttt{TacoxDNA} \cite{suma19_tacox} conversion utility. Usage examples of the visualization software and file conversion facilities are given in \href{https://youtu.be/3zEAnOo9BIo}{supplementary video 4}.

\begin{figure}[H]
    \centering
    % \begin{subfigure}[b]{0.3\textwidth}
    %     \centering
    %     \includegraphics[width=\textwidth]{./img/array}
    %     \caption{\bf{TODO Make array figure, use that here}}
    %     \label{fig:hybrid-array}
    % \end{subfigure}

    \begin{subfigure}[b]{0.49\textwidth}
        \centering
        \includegraphics[width=\textwidth]{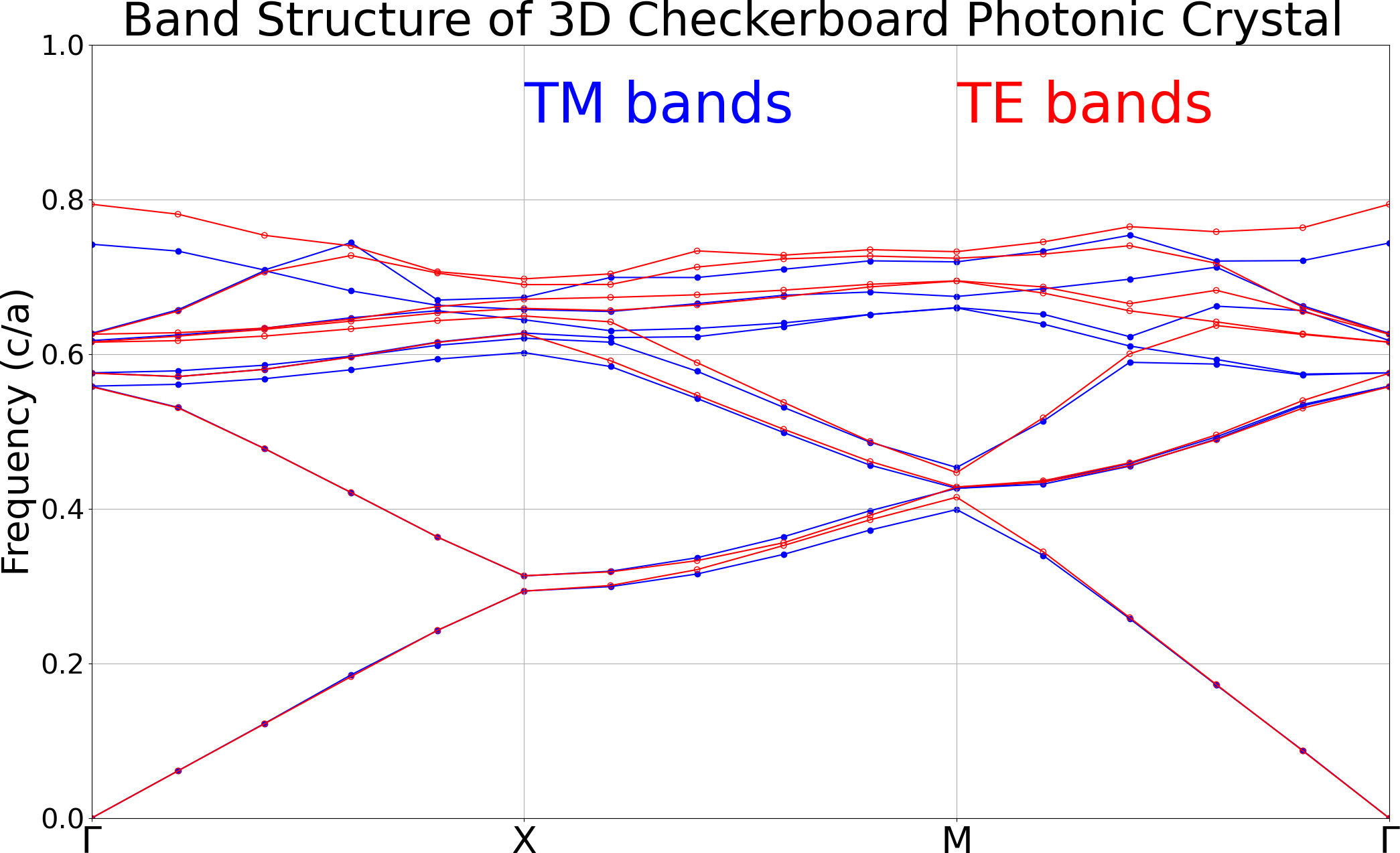}
        \caption{}
        \label{fig:mpb-ex}
    \end{subfigure}
    \begin{subfigure}[b]{0.49\textwidth}
        \centering
        \includegraphics[width=\textwidth]{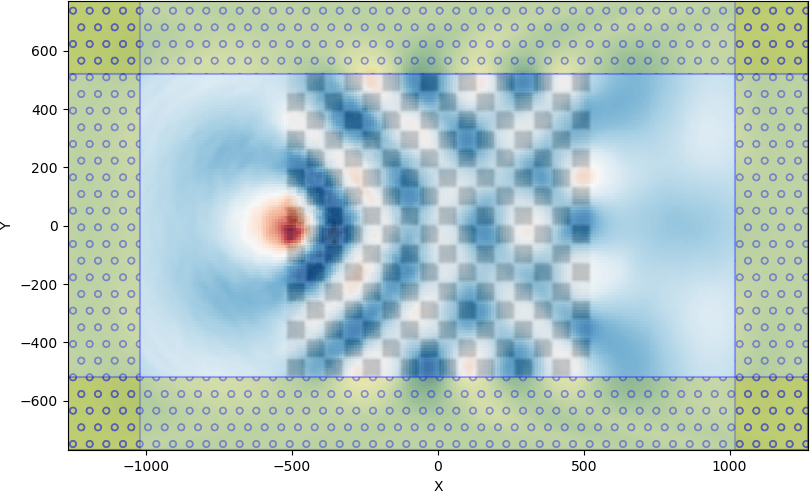}
        \caption{}
        \label{fig:meep-ex}
    \end{subfigure}
       \caption{(a) Band structure of 3D checkerboard photonic crystal composed of 65$nm^3$ cubic $TiO_2$ and $SiO_2$ nanoparticles encased in DNA origami suits. (b) The mode pattern of a \(16\times 16 \times 16\) checkerboard array of $TiO_2$ and $SiO_2$ nanoparticles encased in DNA origami suits illuminated by a continuious source of 500nm wavelength light. Plots created with matplotlib using the MEEP/MPB extension.}
       \label{fig:em-integration}
\end{figure}

While molecular scale simulations can help give confidence that designed structures have the desired shape, they are not useful for modeling macroscopic material properties arising from nanoscale structure. For that, simulation on macroscopic level is needed. As an example of integrating macroscopic scale simulation packages the electromagnetic simulation software packages MIT Photonic Bands, \texttt{MPB}, \cite{johnson01_block_iterat_frequen_domain_method} and MIT Electromagnetic Equation Propagation, \texttt{Meep}, \cite{oskooi10_meep} have been integrated for the simulation of the band structure of photonic crystals and optical response of materials respectively. Using these packages the band structure of a 3D checkerboard photonic crystal made of \ce{TiO2} and \ce{SiO2}, figure \ref{fig:mpb-ex}, and the mode profile of a \(16 \times 16\) cubic array of \ce{TiO2} and \ce{SiO2} arranged in a checkerboard pattern, figure \ref{fig:meep-ex}, have been calculated. A detailed example of the integration with the \texttt{MPB} package is given in \href{https://youtu.be/3zEAnOo9BIo}{supplementary video 7}.

\section{A repository for extensions}
\label{repo}
The ability for users to expand to \texttt{small}'s functionally is done through \emph{extensions}. Extensions can be new models of chemical entities, designed nanostructures that use these models or code to integrate other software into \texttt{small}. This section details the philosophy behind creating extensions and how they are distributed.

Conceptually extensions \texttt{small} borrow from the Unix philosophy, emphasising \emph{doing one thing well} and \emph{writing programs that work together} \cite{mcilroy78_unix_time_sharin_system}. This philosophy makes \texttt{small} modular, allowing users to choose extensions which provide the functionality needed to for their particular use case. This modularity has three benefits; ease of development, ease of use and speed of contribution. Requiring each extension to do only one thing means that extensions can be implemented with relative ease and in few lines of code. Additionally, being self contained makes extensions easier use as users need only to familiarize themselves with the functionality and usage of the specific extensions they are using and not try how to perform specific task using a monolithic software package. This modularity also increases the speed at which users can make contributions and grow the \texttt{small} ecosystem as no central authority needs to review and incorporate extensions into a monolithic code base.

\begin{figure}[H]
    \centering
    % \begin{subfigure}[b]{0.33\textwidth}
    %     \centering
     %     \includegraphics[width=\textwidth]{./img/oxview-triangle}
    %     \caption{}
    %     \label{fig:repo-code}
    % \end{subfigure}%
    \begin{subfigure}[b]{0.5\textwidth}
        \centering
        \includegraphics[width=\textwidth]{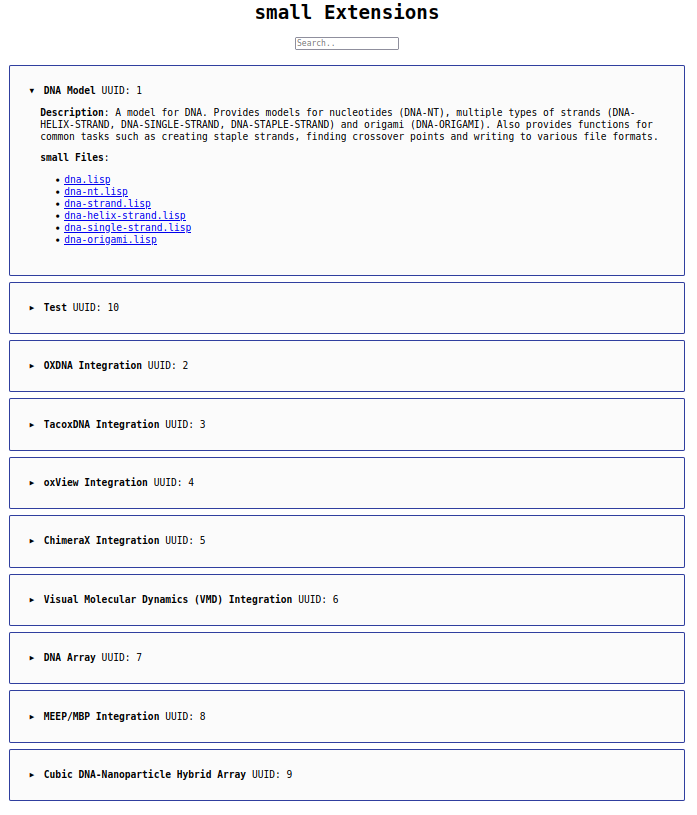}
 %       \label{fig:repo-search}
    \end{subfigure}%
    \begin{subfigure}[b]{0.5\textwidth}
        \centering
        \includegraphics[width=\textwidth]{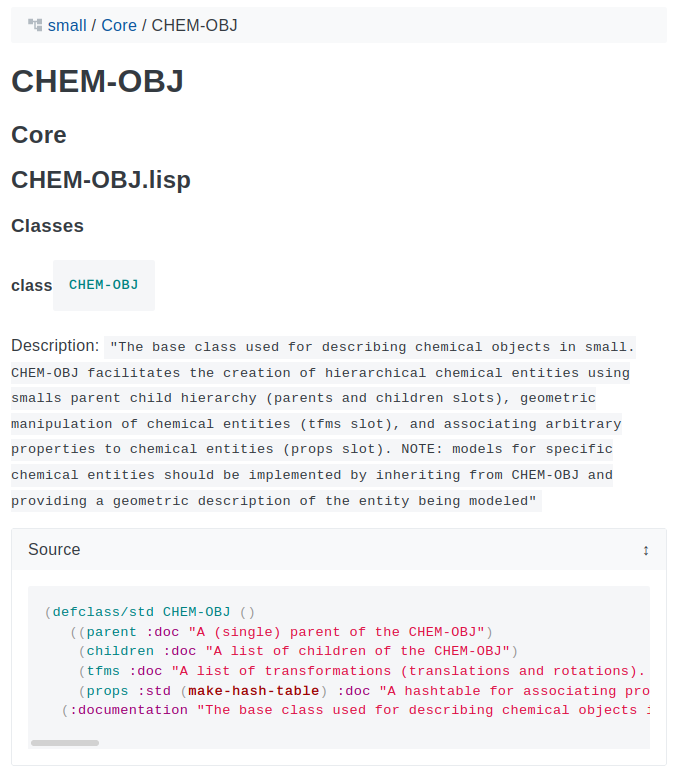}
%        \label{fig:repo-docs}
    \end{subfigure}
       \caption{(a) A view of the repository listing available designs (clicking on a design reveals more detail). (b) An example of the automatically generated documentation for extensions.}
       \label{fig:repo}
\end{figure}

To facilitate the distribution and use of extensions \texttt{small} provides a \emph{repository} and \emph{package manager}. The repository is a central location where extensions to \texttt{small} are stored. Users are able to search the repository for extensions through the web interface as shown in \ref{fig:repo}. In addition to the \texttt{Common Lisp} code that implements the extension repository entries can also store arbitrary information such as experimental protocols for assembling the structure, simulation and experimental results, and articles related the given structure. Documentation for extensions is automatically generated from the docstrings in the \texttt{Common Lisp} code implementing the extension. Additionally users can add their own custom documentation for describing the structure or extension and it's use, an example of the generated documentation is shown in figure \ref{fig:repo}. \texttt{small}'s package manager handles downloading and installing extensions from \texttt{small}'s repository. The use of \texttt{small}'s repository and package manager is detailed in \href{https://youtu.be/8v9u-20fAwI}{supplementary video 8}

\section{Outlook}
\label{sec:org0899b59}
We have introduced an extensible, modular framework for the programmatic design and modeling of arbitrary nanostructures and systems. Within this framework we have used the extension mechanism to create an ICME for DNA nanotechnology capable of designing and modeling 3-dimensional, hierarchical, multi origami structures. This was achieved by providing a model for describing DNA and integrating with the popular DNA molecular dynamics simulation \texttt{oxDNA} and the visualization software \texttt{oxView}, \texttt{ChimeraX} and \texttt{VMD}. Furthermore we have demonstrated the ability to design and model hybrid systems on the micro and macro scale in \texttt{small}'s through the addition of a model for describing cubic nanoparticles and integration with the software packages \texttt{Meep} and \texttt{MPB} for the simulation of the mode profile and band structure of hybrid DNA-nanoparticle structures respectively.

\texttt{small}'s modularity is achieved through the addition of user extensions which, along with designed nanostructures, can be easily uploaded to \texttt{small}'s repository and obtained using it's package manager.

Future work will address adding models of different chemical entities, integrating numerical optimization packages to enable automated structural refinement of nanostructures and integrating with laboratory equipment to allow for the automation of synthesis and elucidation protocols. Furthermore integration with virtual reality visualization software and a natural language interface, where users can describe their desired nanostructure using voice input, are planned.

\section{Code availability}
\label{sec:org68703ec}
\texttt{small} is available on GitHub at \url{https://github.com/DurhamSmith/small}.

\label{bibliographystyle link}
\bibliographystyle{unsrt}

\label{bibliography link}
\bibliography{../../../../org/references}
\end{document}